# Robust conversion of singlet spin order in coupled spin-1/2 pairs by adiabatically switched RF-fields

Andrey N. Pravdivtsev,[a,b,*] Alexey S. Kiryutin,[a,b,*] Alexandra V. Yurkovskaya,[a,b] Hans-Martin Vieth,[a,c] Konstantin L. Ivanov[a,b,**]

[a] *International Tomography Center, Siberian Branch of the Russian Academy of Science, Institutskaya 3A, Novosibirsk, 630090 (Russia)*
[b] *Novosibirsk State University, Pirogova 2, Novosibirsk, 630090 (Russia)*
[c] *Institut für Experimentalphysik, Freie Universität Berlin, Arnimallee 14, Berlin, 14195 (Germany)*

\* These authors have contributed equally

\*\* Corresponding author; tel. +7(383)330-8868, fax +7(383)333-1399, e-mail: Ivanov@tomo.nsc.ru

**Abstract**

We propose a robust and highly efficient NMR technique to create singlet spin order from longitudinal spin magnetization in coupled spin-½ pairs and to perform backward conversion (singlet order)→magnetization. In this method we exploit adiabatic switching of an RF-field in order to drive transitions between the singlet state and the $T_\pm$ triplet states of a spin pair under study. We demonstrate that the method works perfectly for both strongly and weakly coupled spin pairs, providing a conversion efficiency between the singlet spin order and magnetization, which is equal to the theoretical maximum. We anticipate that the proposed technique is useful for generating long-lived singlet order, for preserving spin hyperpolarization and for assessing singlet spin order in nearly equivalent spin pairs in specially designed molecules and in low-field NMR studies.

## I. Introduction

Nuclear singlet states with extended lifetimes [1, 2] are becoming a powerful tool in Nuclear Magnetic Resonance (NMR) spectroscopy. Such Long-Lived spin States (LLSs) can be formed in pairs of spins ½: in such spin pairs the singlet spin state is often long-lived for the reason that it is immune to mutual dipolar relaxation, which typically gives the dominant contribution to spin relaxation rates. LLSs lifetimes, $T_S$, can by far exceed nuclear $T_1$-relaxation times: for instance, for the β-CH$_2$ protons of partially deuterated aromatic amino acids they can be as long at $45 \cdot T_1$ [3, 4]. Extremely long LLS lifetimes have been found in the $^{15}$N-labeled N$_2$O molecule [5]; recently, $T_S$ of the order of 1 hour for a pair of $^{13}$C spins has been reported for specially designed $^{13}$C-labeled molecules [6]. LLSs can be sustained once the nuclear singlet state is an eigen-state (or nearly an eigen-state) of the spin Hamiltonian. Such a condition can be fulfilled by placing the spins at a sufficiently low magnetic field [7], by applying strong RF-excitation (spin-locking) [7] or by using specially designed molecules with "nearly-equivalent" pairs of spins [6]. LLSs create a unique resource for studying slow dynamics processes, notably, for probing slow molecular motions [8, 9], slow diffusion and transport [10-14] and drug-screening [15], when fast $T_1$-relaxation imposes restrictions for the NMR observation window. Another promising application of LLSs is strong non-thermal nuclear spin polarization [16-19]. Such a polarization, often termed hyperpolarization, provides an enormous gain in NMR signal intensity but irreversibly decays due to $T_1$-relaxation. In such a situation LLSs can be a remedy: the precious hyperpolarization stored in the nuclear spin order often has extended lifetimes.

A prerequisite for utilizing singlet spin order in various NMR applications is a robust method for conversion of spin magnetization into the singlet spin order and back, required for creating LLSs and also for observing NMR signals originating from LLSs. There has been a number of techniques developed [1], which provide the desired spin order conversion; however, none of them works efficiently for both weakly and strongly coupled spins. In this context, we would like to mention a technique proposed by some of us [20, 21], which is based on slowly (adiabatically) switching a spin-locking RF-field to enable highly efficient magnetization-to-singlet (M2S) and singlet-to-magnetization (S2M) conversion. The method is based on correlating nuclear spin states in the RF-rotating frame of reference; we have demonstrated that it provides excellent conversion efficiency (equal to the maximal theoretical value [22]) and enables suppression of residual NMR signals originating from spins having no LLSs. By varying the RF-frequency we can select the pathway of spin order conversion, namely, we can selectively induce transitions of the kind $S \leftrightarrow T_+$ or $S \leftrightarrow T_-$. Previously, the method has been applied only to weakly coupled spin pairs, i.e., when the coupling strength, $J$, is much smaller than the difference, $\delta\nu$, in the NMR frequencies. This can be a limitation of the method as experiments on LLSs are often performed on strongly-coupled spin systems, i.e., in the situation where $|J|\sim\delta\nu$ or even $|J| \gg \delta\nu$. With the aim to extend the capabilities of the technique, here we demonstrate that it is operative at an arbitrary relation between $J$ and $\delta\nu$ and provides excellent M2S/S2M conversion. Hence, we show that there is a universal method for robust M2S and S2M conversion, which is efficient for an arbitrary spin-½ pair. We term this technique "*Adiabatic-Passage Spin Order Conversion*", or *APSOC*, without regard to the coupling regime. To demonstrate the power of our method we apply it to a peptide having strongly and weakly coupled spin pairs, namely, $CH_2$-groups. Our results clearly show that the new method is useful for creating singlet spin order and very versatile. Possible applications of the APSOC technique are discussed.

**II. Theory**

Let us describe how APSOC works in a system of two coupled spins upon a perfectly adiabatic RF-field switch. In such a situation the switch becomes reversible. This means that, if we know, for instance, how magnetization is converted into the singlet spin order by an RF-on switch, we also know how the singlet state population is converted by turning off an RF-field: the pathway of spin order conversion is simply reverted upon the RF-off switch. Thus, it is sufficient to correlate the spin states in the rotating frame for zero $\nu_1$ and for "strong" $\nu_1$. For the sake of simplicity, let us discuss the case of nearly equivalent spins, i.e., $|J| \gg |\delta\nu|$, and then generalize the results.

The Hamiltonian of two coupled spins (spin "a" and spin "b") in an external magnetic field $B_0$ directed along the $z$-axis and an oscillating RF-field directed along the $x$-axis in the laboratory frame is as follows (as written in the units of $\hbar$):

$$\widehat{H}^{lf} = -\nu_a \hat{I}_{az} - \nu_b \hat{I}_{bz} - 2\nu_1 \cos(2\pi\nu_{rf}t) \cdot (\hat{I}_{ax} + \hat{I}_{bx}) + J(\hat{\mathbf{I}}_a \cdot \hat{\mathbf{I}}_b) \qquad (1)$$

Here $\hat{\mathbf{I}}_a$ and $\hat{\mathbf{I}}_b$ are the spin operators of the two nuclei under study; $\nu_a$ and $\nu_b$ are the Larmor precession frequencies of the nuclei "a" and "b" at a field $B_0$; $\nu_1$ is the RF-field amplitude. Let us introduce the "center of the spectrum" frequency: $\nu_0 = \frac{(\nu_a+\nu_b)}{2}$ and the difference of Larmor frequencies $\delta\nu = \nu_a - \nu_b$. The offset of the RF-frequency, $\nu_{rf}$, from the center of the spectrum is equal to $\Delta = \nu_0 - \nu_{rf}$. It is convenient to describe the spin dynamics in the frame of reference, which rotates about the z-axis at the frequency $\nu_{rf}$. The Hamiltonian in such a rotating frame is as follows (when the counter-rotating RF-field component is left out):

$$\hat{H}^{rf} = -\Delta_a \hat{I}_{az} - \Delta_b \hat{I}_{bz} - \nu_1(\hat{I}_{ax} + \hat{I}_{bx}) + J(\hat{\mathbf{I}}_a \cdot \hat{\mathbf{I}}_b) \qquad (2)$$

Here $\Delta_a = \nu_a - \nu_{rf}$ and $\Delta_b = \nu_b - \nu_{rf}$ are the offsets of the RF-frequency from the resonance frequencies of spins "a" and "b", respectively. Using the definitions $\Delta$ and $\delta\nu$ we obtain: $\Delta_a = \Delta + \frac{\delta\nu}{2}$ and $\Delta_b = \Delta - \frac{\delta\nu}{2}$. Here we consider both positive and negative $\Delta$. Henceforth, in our analysis we assume that the RF-field strength, $\nu_1(t)$, is a function of time and search for the solution in the simplest case where $\nu_1(t)$ is changed in an adiabatic fashion. It is convenient to present the Hamiltonian as a sum of the main part, $\hat{H}_0$, and the perturbation, $\hat{V}$,

$$\hat{H} = \hat{H}_0 + \hat{V} \qquad (3)$$

For the sake simplicity, let us assume that the spin system is coupled strongly and obtain the results for this specific case and then generalize them for arbitrary relation between $J$ and $\delta\nu$. In such a case the two contributions to $\hat{H}$ are

$$\hat{H}_0 = -\Delta(\hat{I}_{az} + \hat{I}_{bz}) - \nu_1(\hat{I}_{ax} + \hat{I}_{bx}) + J(\hat{\mathbf{I}}_a \cdot \hat{\mathbf{I}}_b), \quad \hat{V} = -\delta\nu(\hat{I}_{az} - \hat{I}_{bz})/2 \qquad (4)$$

In our analysis we stay with $\hat{H}_0$ and correlate the adiabatic levels of the main part at $\nu_1 = 0$ and at $\nu_1 > |J|$. By correlation we mean that the population of the highest level (in energy) at $\nu_1 = 0$ goes to the population of the highest level at $\nu_1 > |J|$; likewise, the population of the second highest state in the absence of the RF-field goes to the population of the second highest state in the presence of the RF-field and so on. Importantly, while performing the correlation of states we always stay in the RF-rotating frame (even when $\nu_1$ is zero) in order to remove all rapidly oscillating terms in the Hamiltonian, compare eqs. (1) and (2). The perturbation $\hat{V}$ repels the levels of $\hat{H}_0$, when they tend to cross. This situation is known as a Level Anti-Crossing (LAC). The solution of $\hat{H}_0$ can be obtained for an arbitrary RF-field strength; a "good" basis for this Hamiltonian is the singlet-triplet basis. However, one should note that, in general, the quantization axis for the triplet states is neither z, nor x: this axis is parallel to the effective field experienced by the spins. Such a frame is termed "tilted" frame and $\hat{H}_0$ in this frame is written as

$$\hat{H}_{0,tf} = -\nu_{eff}(\hat{I}_{az} + \hat{I}_{bz}) + J(\hat{\mathbf{I}}_a \cdot \hat{\mathbf{I}}_b) \qquad (5)$$

where $\mathbf{v}_{eff} = (\nu_1, 0, \Delta)$, so that the effective field is the following vector: $\mathbf{B}_{eff} = -2\pi \frac{\mathbf{v}_{eff}}{\gamma}$ (here $\gamma$ is the nuclear gyromagnetic ratio).

Let us first consider what happens when $\nu_1 = 0$, i.e., when the RF-field is off. The main Hamiltonian, $\hat{H}_{0,tf}$, has four eigen-states, $\{S, T_+, T_0, T_-\}$, with the following energies:

$$E(T_+) = -\Delta + \frac{J}{4}, \quad E(T_-) = \Delta + \frac{J}{4}, \quad E(T_0) = \frac{J}{4}, \quad E(S) = -\frac{3J}{4} \qquad (6)$$

The order, in which these states are grouped, depends on the signs of $J$ and $\Delta$; the sign of $\Delta$ is set by varying the RF-frequency, see **Figure 1a** and **1b**. When $|\Delta| > |J|$ the $S$ and $T_0$ states are always the second and third highest states in energy. As we show below, this is not what we want: APSOC only works in the opposite case, i.e., when $|J| > |\Delta|$. In this situation the states are grouped as shown in **Figure 1a** and **1b** (shown for $J > 0$). Thus, the lowest state is always the singlet state; the three triplet states are grouped according to the sign of $\Delta$. It is important to note that at $v_1 = 0$ the triplet states are those defined for the quantization axis parallel to the $z$-axis.

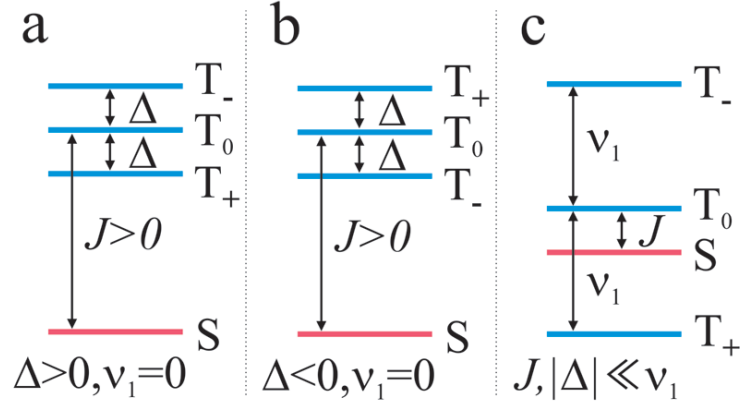

**Figure 1**. Energy levels of a strongly coupled two-spin system (i.e., $J \gg \delta v$) in the RF-rotating frame in the absence of an RF-field (a and b) and in the presence of a strong RF-field (c). In (a) $\Delta > 0$ and the second lowest state is the $T_+$ state; in (b) $\Delta < 0$ and the second lowest state is the $T_-$ state. In (c) we assume $v_1 \gg J, |\Delta|$. The spin states $S, T_0, T_\pm$ are the eigen-states in the "tilted" frame where the $z$-axis is parallel to the effective field vector, $\mathbf{B}_{eff}$. Here $|\Delta| < |J|$ and $J > 0$.

When the RF-field is on and it is sufficiently strong, $v_1 \gg |\Delta|, |J|$, we obtain:

$$E(T_+) = -v_1 + \frac{J}{4}, \quad E(T_-) = v_1 + \frac{J}{4}, \quad E(T_0) = \frac{J}{4}, \quad E(S) = -\frac{3J}{4} \qquad (7)$$

Thus, the states are grouped as shown in **Figure 1c**; namely, the second lowest state in energy is the $S$ state.

Consequently, upon increasing the RF-field strength the singlet state should be correlated with the $T_+$ state (case $\Delta > 0$) or $T_-$ state (case $\Delta < 0$). At thermal equilibrium the $T_+$ state is overpopulated, whereas the $T_-$ state is underpopulated; thus, the two kinds of conversion provide an overpopulated or underpopulated singlet state. The method works *only* when $|\Delta| < |J|$; otherwise, we do not change the order of the adiabatic states and only have the correlation of the kind $S \leftrightarrow S$.

In the discussion we so far omitted the $\hat{V}$ term. In fact, this term is very important. The reason is that at a particular $v_1$ strength, namely, when $v_{eff} = \sqrt{v_1^2 + \Delta^2}$ is equal to $|J|$ we obtain an $S$-$T_+$ (or $S$-$T_-$) crossing for the main Hamiltonian, $\hat{H}_{0,tf}$, see **Figure 2**. Consequently, in the absence of the perturbation there is no $S \leftrightarrow T_+$ or $S \leftrightarrow T_-$ conversion: instead the correlation of states is $S \leftrightarrow S$, $T_0 \leftrightarrow T_0$ and $T_\pm \leftrightarrow T_\pm$, i.e., the APSOC method does not work. So, the method works only because at the crossing we have the $\hat{V}$ term, which mixes the crossing levels and makes an LAC out of the crossing. The time of passage through the crossing thus has to be of the order of $1/\delta v$ and does not have to be very precisely controlled (when the relaxation time is much greater than $1/\delta v$). It is worth noting that when the sign of $J$ is negative the conversion pathways change as follows: it is $S \leftrightarrow T_-$ conversion (case $\Delta > 0$) and $S \leftrightarrow T_+$ conversion (case $\Delta < 0$). This change of the APSOC pathway is of no principal importance, since APSOC works in the same way as for $J > 0$.

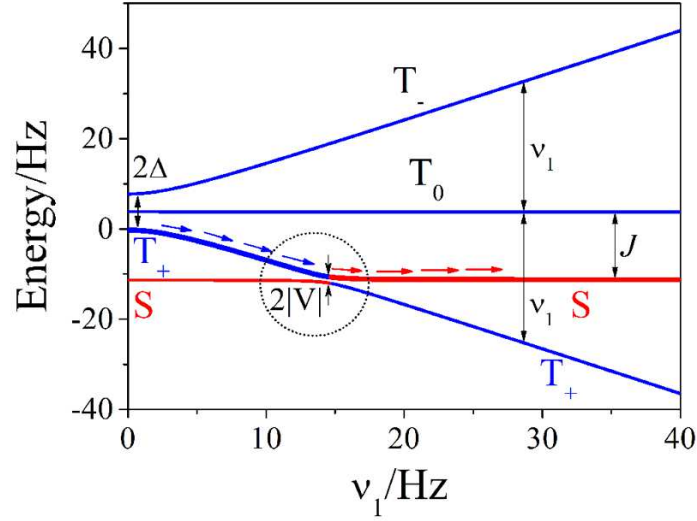

**Figure 2**. Adiabatic correlation of states in the APSOC method. Energy levels of the four spin states of a strongly coupled spin pair are shown as functions of $v_1$ for the following parameters: $J = 15$ Hz, $\Delta = 4$ Hz, $\delta v = 2$ Hz. In this example we have correlation of the $T_+$ state at $v_1 = 0$ with the singlet state, $S$, at $v_1 \gg J$; the corresponding adiabatic level, which is the second lowest level in energy, is highlighted and the spin order conversion pathway is indicated by arrows. This level has an LAC (indicated by the circle) with the lowest energy level; a prerequisite for the desired spin order conversion, $S \leftrightarrow T_+$, is adabatic passage through this LAC. The relevant splitting between the energy levels is indicated at $v_1 = 0$ and strong $v_1$ and also at the LAC. When $\Delta < 0$ the energy level diagram looks the same except that the states $T_+$ and $T_-$ at $v_1 = 0$ are exchanged.

Interestingly, our method also works for an arbitrary relation between $J$ and $\delta v$. The reason for this is that at $v_1 = 0$ two of the spin eigen-states are $T_+$ and $T_-$ (the other two states are given by superposition of the states $S$ and $T_0$) whereas at $v_1 \gg |J|, |\Delta|$ one of the eigen-states is the singlet state. In addition, it is possible to set the off-set $\Delta$ such that $T_+$ (or $T_-$) is the second lowest state in energy (in the RF-rotating frame) at $v_1 = 0$. Therefore, by an appropriate choice of the off-set, $\Delta$, we can still perform conversion of the kind $S \leftrightarrow T_-$ or $S \leftrightarrow T_+$. In the general case, when the relation between $J$ and $\delta v$ is arbitrary, the value of $\Delta$ should be set as follows for APSOC (for positive $J$):

$$|\Delta| < \frac{1}{2}\sqrt{\delta v^2 + J^2} + \frac{J}{2} \qquad (8)$$

Thus, APSOC works for an arbitrarily coupled pair. In order to make easy the search for the optimal RF-field parameters (i.e., $v_{rf}$, maximal $v_1$ value and RF-field switching time) for APSOC (for an arbitrary relation between $J$ and $\delta v$) we developed software, which is available online [23]. Details of optimization of the RF-field switches in the APSOC are presented in Appendix A.

An important point here is the performance of our method with respect to the M2S and S2M conversion efficiency. Assuming that relaxation during RF-excitation equalizes the populations of the three triplet states but does not affect the singlet state, in APSOC we expect that after the two conversion stages the spin magnetization, $M$, equal to $\frac{2}{3}$ of the initial magnetization, $M_0$, remains. However, experimentally detected $M$ can be higher than $\frac{2}{3}M_0$ when the triplet relaxation is incomplete (in the limiting case where the RF-excitation period is shorter than all relaxation times in the spin system we obtain $M = M_0$ due to reversibility of adiabatic transitions), see Appendix B for details. The $M = \frac{2}{3}M_0$ value is the theoretical upper limit for the M2S-S2M conversion efficiency [22]. Of course, as the singlet maintenance time increases the resulting magnetization decreases as well due to singlet-order relaxation. Hereafter the value $M/M_0$ is denoted as $\varepsilon$.

### III. Results

To demonstrate the utility of the APSOC method we use the experimental protocol shown in **Figure 3**. The RF-field frequency for the $S \leftrightarrow T_-$ or $S \leftrightarrow T_+$ conversion is chosen as shown in **Figure 3a** for a strongly-coupled spin pair. The protocol comprises four stages, see **Figure 3b**: (M2S conversion)-(singlet order maintenance)-(S2M conversion)-(NMR detection). For the M2S/S2M conversion we exploit APSOC and set Δ in order to obtain spin order conversion of the kind $S \leftrightarrow T_-$ or $S \leftrightarrow T_+$. For singlet maintenance we use a spin-locking field in order to suppress singlet-triplet mixing. In principle, for strongly coupled spin pairs, $J \gg \delta\nu$, singlet maintenance is possible even without introducing a spin-locking field. Additionally, we perform a comparison of APSOC and the so-called Spin-Locking Induced Crossing (SLIC) method [24], which is a technique for generating LLS in pairs of nearly equivalent spins. Details of the SLIC method are given in Appendix C.

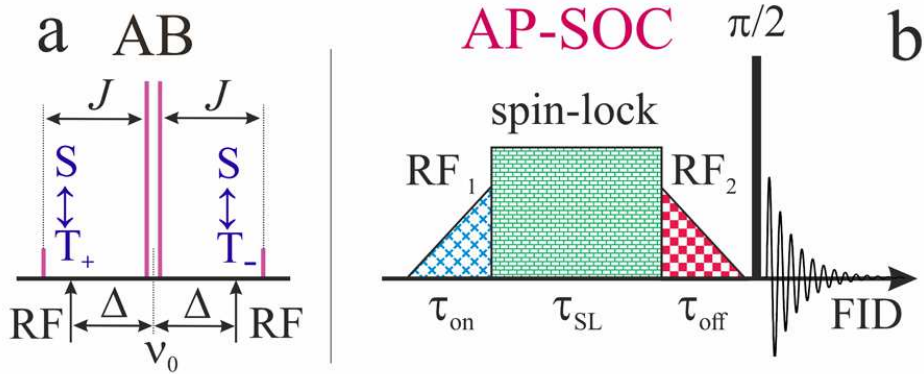

**Figure 3**. (a) Setting of the RF-frequency for APSOC and (b) experimental protocol used for generating, sustaining and observing LLSs. In (a) the off-sets, $\pm\Delta$, of the RF-frequency from $\nu_0$ are shown, which provide $S \leftrightarrow T_-$ or $S \leftrightarrow T_+$ conversion (here, for a strongly coupled pair $|\Delta| < J$). The protocol shown in (b) comprises 4 stages. In stage 1 magnetization is converted into the singlet state (by APSOC we perform conversion of the kind $T_+ \rightarrow S$ or $T_- \rightarrow S$) by the RF$_1$-field, which is adiabatically turned on during the time period $\tau_{on}$. In stage 2 the singlet state is sustained during a variable time interval $\tau_{SL}$ (by using spin locking, or without spin-locking when $J \gg \delta\nu$). In stage 3 the singlet state is converted back into $z$-magnetization by the RF$_2$-field, which is adiabatically turned off during the time period $\tau_{off}$. Finally, in stage 4 the NMR spectrum is taken by using a $\pi/2$-pulse. The frequency of the RF$_1$- and RF$_2$-fields is equal to $(\nu_0 \pm \Delta)$.

Experiments were performed for aqueous solutions of a dipeptide, H-Cys-Gly-OH (hereafter, Cys-Gly), at 400 MHz and 700 MHz NMR spectrometers. LLSs are created for the strongly coupled α-CH$_2$ protons of the Gly-residue ($J = 18$ Hz, $\delta\nu$ is 8.6 Hz at 400 MHz and 15 Hz at 700 MHz) and the weakly coupled β-CH$_2$ protons of the Cys-residue. The sample contains 30 mM of Cys-Gly (Bachem, G-3755) and 10 mM of EDTA at pH* 12.0 in D$_2$O. The sample was bubbled by N$_2$ for 10 minutes to remove the dissolved oxygen. EDTA was used to chelate paramagnetic ions. pH* indicates uncorrected reading from a pH-meter calibrated in H$_2$O.

Typical NMR spectra obtained using the APSOC method are shown in **Figure 4a** for different singlet maintenance times, $\tau_{SL}$: at long $\tau_{SL}$ of about 60 s the NMR signals are still visible, clearly indicating the presence of an LLS. The full $\tau_{SL}$ dependence, see Appendix C, allowed us to find the singlet order lifetime $T_S$, which equals $21 \cdot T_1$ at 700 MHz. Thus, we indeed create a long-lived spin order, i.e., singlet order, in the spin pair under study.

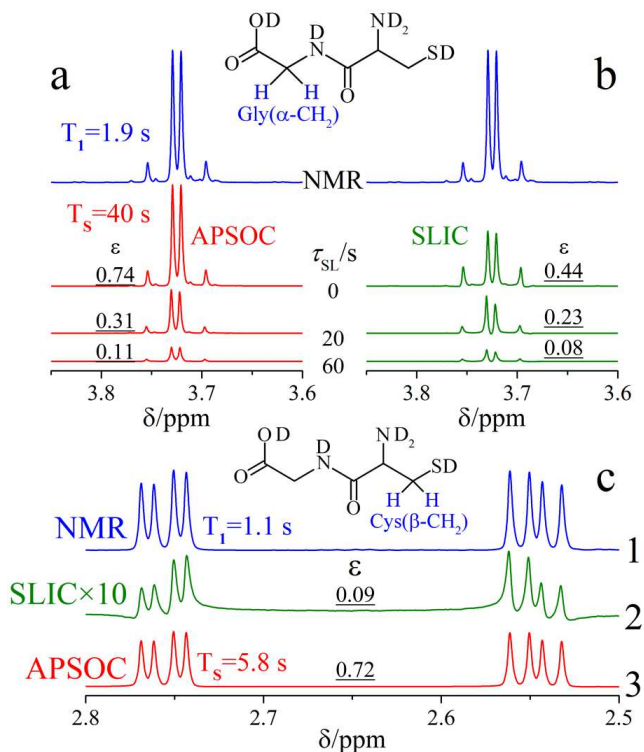

**Figure 4**. (top) 700 MHz $^1$H-NMR spectrum of Gly($\alpha$-CH$_2$) protons of Cys-Gly and APSOC (a) and SLIC (b) spectra for different duration, $\tau_{SL}$, of the spin-locking period. (c) NMR, APSOC and SLIC spectra of the Cys($\beta$-CH$_2$) protons of Cys-Gly. The signal intensities, $\varepsilon$, given in the units of $M_0$, are shown on the spectra by underlined numbers. $T_S$ and $T_1$ times for each pair are indicated. Experimental APSOC and SLIC parameters are given in Appendix C.

Now let us discuss the performance of our spin order conversion method, as demonstrated in **Figure 4b**. One can clearly see that after the M2S-S2M conversion the NMR signal remains strong as compared to the initial thermal signal $M_0$. Namely, the signal is $0.74 \cdot M_0$ at short $\tau_{SL}$ and goes to $\frac{M_0}{2}$ at $\tau_{SL} \approx 3$ s, i.e., at LLS maintenance times, which are longer than $T_1$. When the $\tau_{SL}$-dependence is approximated as a sum of two exponential functions, $M(\tau_S) = A_T \exp(-\tau_{SL}/T_T) + A_S \exp(-\tau_{SL}/T_S)$ (fast and slow exponent, $T_T \approx T_1$), the weight of the slow component, $A_S$, is about $0.51 \cdot M_0$. Thus, our method indeed provides excellent M2S-S2M conversion efficiency. In general, when it becomes possible to reduce the RF-switching time so that $(\tau_{on} + \tau_{off}) \ll T_1$ (in our example, $\tau_{on} + \tau_{off} \approx T_1/3$) we expect that the weight of the slow component reaches $\frac{2}{3} M_0$. Here we attribute the loss of the slow component at $\tau_{SL} \to 0$ to relaxation during the RF-on/RF-off periods. Additionally, we performed SLIC experiments for the same spin pair as described in Appendix C, which provide a similar $\tau_{SL}$ dependence; however, the conversion efficiency in the SLIC case is systematically lower, see **Figure 4c**. For instance, at short $\tau_{SL}$ we obtain $M < \frac{M_0}{2}$ in the SLIC case.

Additionally, we studied the influence of the spin-locking field on the singlet maintenance efficiency. Using such a spin-locking field is not necessary when $J \gg \delta\nu$, so that the singlet state is almost an eigen-state of the Hamiltonian $\widehat{H}$. However, when $J \sim \delta\nu$ spin-locking makes the LLS lifetime longer: at 700 MHz $T_S$ increases from 6 s to 40 s, see Appendix D.

As mentioned above, our method works for both weakly and strongly coupled spin pairs. In this work, the APSOC method was also applied to the weakly-coupled β-CH$_2$ protons of the Cys residue. At $\tau_{SL} \to 0$ we obtain $M = 0.72 \cdot M_0$; the $\tau_{SL}$ dependence shows that we clearly generate an LLS ($T_S = 5.8$ s, which is about $4.8 \cdot T_1$). The SLIC method is not directly applicable to this system, because, literally, there is no LAC at $\nu_1 \approx |J|$ in this case. After a modification the SLIC method can be nonetheless used, but its performance is much lower than for APSOC, see Appendix C.

An important issue for our method is the setting of the $\tau_{on}$ and $\tau_{off}$ times and the maximal $\nu_1$-value of the switched RF-fields. When optimizing these parameters we keep in mind that the adiabaticity condition implies that the rates, at which the eigen-states of $\widehat{H}$ change with time, given by the expression $\langle i|\frac{d}{dt}|j\rangle$, are much smaller than the instantaneous frequencies of coherent spin evolution, $\omega_{ij}(t) = (E_i - E_j)$ (here we introduce the solution of the eigen-problem of the Hamiltonian as $\widehat{H}(t)|i\rangle = E_i(t)|i\rangle$). Here we do not optimize the exact $\nu_1(t)$ time profile and only vary the maximal RF-field strength $\nu_1^{max}$ and the switching time $\tau_{on}$ (or $\tau_{off}$) by running numerical calculations. Examples of such optimization are shown in Appendix A for a pair of strongly-coupled protons and a pair of nearly equivalent $^{13}$C nuclei. Once $\tau_{on}$ and $\tau_{off}$ are above a certain threshold value (minimal time compatible with adiabatic variation of the Hamiltonian) there is no benefit to increase them further, since spin relaxation during the pulses reduces the spin order. The Δ parameter can be set according to Eq. (8). Once these conditions are fulfilled the APSOC performance is insensitive to variation of the RF-field parameters.

One should also note that the RF-field switches between stages 1,2,3, see Figure 3, result in non-adiabatic changes of the spin Hamiltonian. However, these switches do not affect the singlet state population (because the singlet is nearly an eigen-state of the Hamiltonian in all cases) and do not disturb the singlet spin order.

**IV. Conclusions**

Thus, a general method for generating and observing singlet spin order is proposed. The method makes efficient use of adiabatically switched RF-fields, which perform the M2S and S2M conversion of the kind $S \leftrightarrow T_+$ and $S \leftrightarrow T_-$. APSOC works for strongly coupled spin pairs as well as for weakly coupled spin pairs. The technique is simple in use and requires setting two parameters, the frequency of the switched RF-field and the ramp of RF-field switching, whereas the $\tau_{on}$ and $\tau_{off}$ times, as well as $\nu_1^{max}$, do not need to be carefully controlled. Thus, in contrast to methods, which exploit spin coherences, by using adiabatic passage we do not need to control the experimental timing. The switched RF-field thus works as an NMR "pulse", which induces spin order conversion of the kind $S \leftrightarrow T_+$ or $S \leftrightarrow T_-$. Furthermore, by varying the frequency of the switched RF-fields, one can change the "phase" of the APSOC "pulses" and arrange a pseudo "phase cycle" [20]. As will be shown in a separate publication, such a filtering procedure suppresses residual background signals, leaving only the signals of spin pairs under investigation. This makes our method very useful for selecting signals of a particular spin pair and for analyzing LLSs in a situation where the NMR spectrum is crowded. Thus, our general method for generating and observing singlet spin order creates a new powerful resource for studying LLSs in protein macromolecules and RNA/DNA fragments.

We anticipate that our method is useful in many other situations as well. For instance, it can be applied to systems with more than two spins, e.g., to three-spin systems [20]. We also expect that the method works not only for protons, but for other nuclei as well. APSOC can be a method of choice to generate and observe LLSs in pairs of nearly equivalent spins, which are found in low-field experiments [7, 25] and specially designed molecules [6, 12, 13, 18]. The ease of our method (as there is no need to set carefully the RF-field strength and duration of the RF-pulse), the possibility of using APSOC for spin pairs with an arbitrary relation between $J$ and $\delta\nu$ creates a new resource for generating and observing LLSs. One more topical application is preserving and manipulating spin hyperpolarization [16-19]. In particular, our method can be used in Para-Hydrogen Induced Polarization (PHIP) experiments [26, 27]: when PHIP is generated in a pair of strongly coupled protons (e.g., at low magnetic fields) there is a need to convert the spin order into observable spin magnetization. We expect that the proposed technique perfectly fits to this experimental need. Likewise, APSOC is compatible with other hyperpolarization methods notably, with dynamic nuclear polarization: presently, the most general hyperpolarization technique. APSOC can be used to overcome the severe limitation imposed by $T_1$-relaxation, which is a stumbling block for many applications of hyperpolarization in NMR spectroscopy and imaging. Last but not least, our method can be used to generate "entangled" quantum states, which are in the heart of quantum computing and quantum information processing [28, 29]

**Acknowledgements**

Experimental studies are supported by the Russian Science Foundation (project No. 15-13-20035); we acknowledge the Russian Foundation for Basic Research (projects No. 14-03-00397 and 15-33-20716) and FASO of Russia for providing access to NMR facilities (project No. 0333-2014-0001).

**APPENDIX A. APSOC performance and optimization**

To find out what are the conditions for adiabatic RF-field switches in APSOC we have performed numerical calculations of the singlet order conversion efficiency.

The thermal density matrix of a two spin-system is as follows:

$$\hat{\rho}_0 = \frac{\hat{E}}{4} + \xi(\hat{I}_{az} + \hat{I}_{bz}) = \frac{\hat{E}}{4} + \xi(\hat{\rho}_{T_+} - \hat{\rho}_{T_-}) \quad (A1)$$

Here $\xi$ is the Boltzmann factor, $\hat{I}_{1z}$ and $\hat{I}_{2z}$ are the z-projections of spins "a" and "b"; consequently the matrices $\xi\hat{\rho}_{T_+}$ and $\xi\hat{\rho}_{T_-}$ account for the equilibrium populations of the $|T_+\rangle$ and $|T_-\rangle$ states. We also introduce the density matrix of the $|S\rangle$ state, $\hat{\rho}_S$. When necessary, the matrices $\hat{\rho}_S$, $\hat{\rho}_{T_-}$ and $\hat{\rho}_{T_+}$ can be writen in the spin-operator form:

$$\hat{\rho}_{T_+} = \frac{\hat{E}}{4} + \frac{\hat{I}_{az}}{2} + \frac{\hat{I}_{bz}}{2} + \hat{I}_{az}\hat{I}_{bz} \quad (A2a)$$

$$\hat{\rho}_{T_-} = \frac{\hat{E}}{4} - \frac{\hat{I}_{az}}{2} - \frac{\hat{I}_{bz}}{2} + \hat{I}_{az}\hat{I}_{bz} \quad (A2b)$$

$$\hat{\rho}_S = \frac{\hat{E}}{4} - (\hat{\mathbf{I}}_a \cdot \hat{\mathbf{I}}_b) \quad (A2c)$$

In APSOC transfer of population of the $|T_+\rangle$ or $|T_-\rangle$ states into population of the $|S\rangle$ state (and vice versa) occurs, depending on the RF-frequency. When we apply the adiabatic $T_+ \to S$ and $T_- \to S$ pulses, the density matrix changes as follows:

$$\hat{\rho}_0 \xrightarrow{T_+ \to S} \frac{\hat{E}}{4} + \xi\hat{\rho}_S + \hat{\rho}_+ \quad (A3a)$$

$$\hat{\rho}_0 \xrightarrow{T_- \to S} \frac{\hat{E}}{4} - \xi\hat{\rho}_S + \hat{\rho}_- \quad (A3b)$$

Here the density matrices $\hat{\rho}_+$ and $\hat{\rho}_-$ stand for the residual triplet spin order; both matrices are "orthogonal" to $\hat{\rho}_S$: $\text{Tr}\{\hat{\rho}_+\hat{\rho}_S\} = \text{Tr}\{\hat{\rho}_-\hat{\rho}_S\} = 0$. From Eqs. (A3) we compute the following singlet-state populations:

$$n(S) = \frac{1}{4} + \xi, \quad \text{for} \quad T_+ \to S \quad (A4a)$$

$$n(S) = \frac{1}{4} - \xi, \quad \text{for} \quad T_- \to S \quad (A4b)$$

In the general case, i.e., when adiabaticity is not perfectly fulfilled, the resulting density matrix differs from Eq. (A3). In such a situation the simple relations (A3) do not hold and the final density matrix is equal to $\hat{\rho}_{fin} = \hat{\hat{Q}}\hat{\rho}_0$, where $\hat{\hat{Q}}$ is the super-operator describing the spin evolution; the resulting singlet-state population is equal to

$$n_S = \text{Tr}\{\hat{\rho}_S \cdot \hat{\hat{Q}}\hat{\rho}_0\} = \frac{1}{4} + \xi \cdot \text{Tr}\{\hat{\rho}_S \cdot \hat{\hat{Q}}[\hat{\rho}_{T_+} - \hat{\rho}_{T_-}]\} \quad (A5)$$

To make a quantitative account for imperfect adiabaticity we introduce the following parameter, which characterizes the singlet order conversion efficiency:

$$\eta = \frac{n_S - \frac{1}{4}}{\xi} = \text{Tr}\{\hat{\rho}_S \cdot \hat{\hat{Q}}[\hat{\rho}_{T_+} - \hat{\rho}_{T_-}]\} \quad (A6)$$

Thus $\eta$ is the measure of singlet order conversion efficiency for an arbitratry M2S conversion method. The $\eta$ parameter changes in the range from $-1$ to $+1$. Specifically, for population transfer between the $|T_-\rangle$ and states we obtain $\eta = -1$, while for the $T_+ \to S$ conversion we obtain $\eta = +1$; when $\eta = 0$ the singlet state is neither overpopulated nor underpopulated, i.e., $n_S = \frac{1}{4}$. Hence the absolute value of the $\eta$ parameter should be maximized in order to find the optimal conditions for the M2S/S2M conversion. Here we performed optimization only for the M2S conversion because in the case of adiabatic transitions the S2M conversion proceeds in the same way. For the sake of simplicity, here we did not consider spin relaxation. In a simple way, the relaxation effects can be taken into account by optimizing the RF-switching profile under the constraint that the $\tau_{on}$ time is shorter than the relaxation times of the spin system.

We investigated how the absolute value of the $\eta$ parameter changes upon variation of the $J$ and $\delta\nu$ parameters, as well as the switching time, $\tau_{on}$, and the switching profile.

In **Figure 1A** we show the optimal APSOC efficiency for a pair of protons as a function of $\tau_{on}$. In the calculation we used linear profile and exponential profiles (the results for other profiles can be obtained and analyzed in the same way), which are optimized: for the linear profile the $v_1^{max}$ value is optimized, for the exponential profile two parameters, $v_1^{max}$ and $k$ (characterizing the ramp of the RF-on field), are optimized. It is clearly seen that at long $\tau_{on}$ times the APSOC efficiency approaches unity, whereas at short $\tau_{on}$ it is zero because the conditions for adiabatic variation of the Hamiltonian are not fulfilled. At $\delta\nu \gg J$, i.e., for a weakly-coupled spin system the time compatible with adiabatic variation of the Hamiltonian is given by $1/J$. In the opposite case, $\delta\nu \ll J$, of a strongly-coupled spin system the switching time compatible with adiabatic variation of the Hamiltonian is given by $1/\delta\nu$. "Adiabatic variation" implies that the Hamiltonian is varied slowly; however, one can see that the $\tau_{on}$ times providing the APSOC efficiency of already 0.5 are not that long: they are of the order of $1/J$ (in a weakly-coupled spin pair) or $1/\delta\nu$ (in a strongly-coupled spin pair).

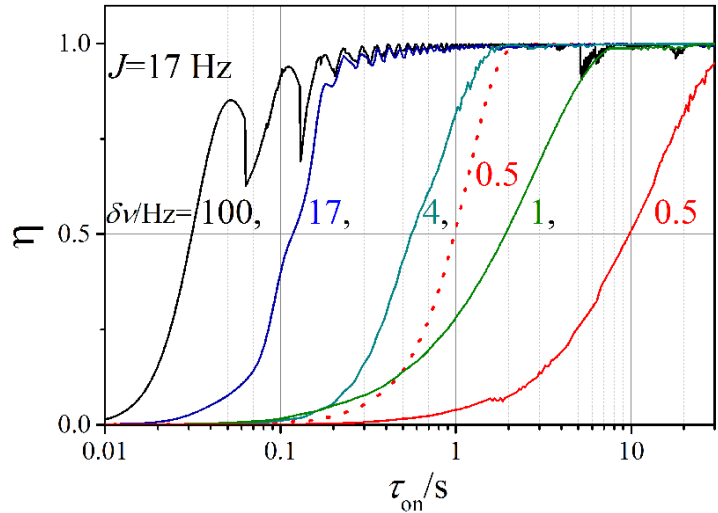

**Figure 1A**. Calculated dependence of $|\eta|$ on the $\tau_{on}$ time of RF-field switching. The APSOC efficiency is measured as described by Eq. (A6). The calculation is done without taking relaxation processes into account. Here $J = 17$ Hz, $\Delta = 10$ Hz, $\delta\nu$ is varied. The RF-field switching on profiles are linear $v_1(t) = v_1^{max}(t/\tau_{on})$ (solid lines) or exponential $v_1(t) = v_1^{max}\frac{1-\exp(-kt)}{1-\exp(-k\tau_{on})}$ (dotted line); here $0 < t < \tau_{on}$. The parameters of the field profile (the $v_1^{max}$ value for the linear profile; the $v_1^{max}$ value and $k$ for the exponential profile) are optimized to obtain the maximal APSOC efficiency for the given spin-system and $\tau_{on}$ and $\Delta$ parameters.

We also evaluated the APSOC performance for a specific system of two nearly-equivalent $^{13}$C spins in a naphthalene derivative, which have an extremely long LLS lifetime [6]. The results are shown in **Figure 2A**. It is readily seen that APSOC provides excellent $T_+ \to S$ conversion efficiency when the $\tau_{on}$ times are about 1 s or longer. Thus, we anticipate that the APSOC method is useful for generating and observing LLSs in such a specifically designed molecule with nearly equivalent spins.

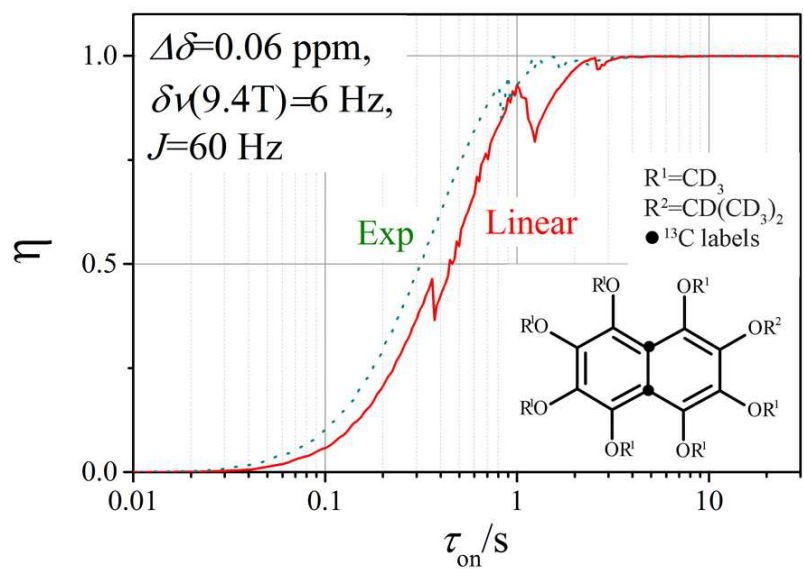

**Figure 2A**. Calculated dependence of $|\eta|$ on the $\tau_{on}$ time of RF-field switching. The APSOC efficiency is measured as described by Eq. (A6). The calculation is done without taking into account relaxation processes and considers two $^{13}$C nuclei of a naphthalene derivative that are indicated by black dots in the Figure. The $^{13}$C spin pair has a long-lived state with a lifetime of ~1 hour [6]. Parameters of the spin system: chemical shift difference is 0.06 ppm, $\delta v(9.4\ \text{T}) = 6$ Hz, $J = 60$ Hz, $\Delta = 10$ Hz. Parameters of the RF-field profile (the $v_1^{max}$ and $k$ values) are optimized.

## APPENDIX B. M2S/S2M conversion efficiency in APSOC

Let us derive the magnetization value after two M2S/S2M conversion stages.

For spins having small thermal polarization we obtain the following populations of the spin states:

$$n(T_+) = \frac{1}{4}(1+P), \quad n(T_-) = \frac{1}{4}(1-P), \quad n(T_0) = n(S) = \frac{1}{4} \tag{B1}$$

and the net spin magnetization is equal to $M_0 = n(T_+) - n(T_-) = \frac{P}{2}$. Here $P$ stands for the population differences between the spin states at equilibrium conditions. Let us assume that spin relaxation during the M2S/S2M conversion step is negligible and assume that we have $S \leftrightarrow T_+$ conversion (the results for $S \leftrightarrow T_-$ conversion are exactly the same). In this situation after the M2S conversion the population of the $S$-state is equal to $\frac{1}{4}(1+P)$. When the singlet maintenance time, $\tau_{SL}$, is longer than the $T_1$-relaxation time but shorter than $T_{LLS}$ the populations of the three triplet states are rapidly equalized and become equal to $\langle n(T) \rangle = \frac{1}{4} - \frac{P}{12}$. After the S2M conversion step the state populations are

$$n(T_+) = \frac{1}{4}(1+P), \quad n(T_-) = n(T_0) = n(S) = \langle n(T) \rangle = \frac{1}{4} - \frac{P}{12} \tag{B2}$$

Consequently, the spin magnetization becomes

$$M = n(T_+) - n(T_-) = \frac{P}{3} = \frac{2}{3}M_0 \tag{B3}$$

Thus, $\frac{1}{3}$ of the initial magnetization is irreversibly lost. Previously it has been shown [22] that such losses are inevitable, i.e., the M2S-S2M conversion efficiency is smaller than 1. When $\tau_{SL} \ll T_1, T_S$ we obtain $M = M_0$ because adiabatic transitions are reversible: the RF$_1$-field redistributes the spin state populations but the RF$_2$-field returns them back to the initial values.

When spin relaxation during the switches comes into play it reduces the resulting $M$ value because during the switch the singlet state is not an eigen-state of the Hamiltonian, with the consequence that all spin states are affected by T$_1$- and T$_2$-relaxation. This results in the loss of singlet spin order, i.e., in reduction of the measured $M$ value.

**APPENDIX C. Comparison of APSOC and SLIC**

In the main text we argue that APSOC is advantageous as compared to the SLIC technique. Here we present further detail of this comparison.

SLIC is dealing with spin order conversion in strongly-coupled spin pairs [24]. The SLIC protocol is presented in **Figure 1C**. In this protocol, first the transverse $y$-magnetization is formed by a $\left(\frac{\pi}{2}\right)_x$-pulse. Then an RF-field is applied during a time period $\tau_{SLIC}$ along the $y$-axis of the rotating frame. The RF-frequency should be equal to $\nu_0$, see **Figure 1Ca**. When the RF-field amplitude matches the $J$ value, it induces M2S transitions, i.e., coherent spin order conversion driven by $\delta\nu$. After $\tau_{SLIC}$ equal to half-period of the coherent evolution the conversion efficiency is maximal. After that singlet-spin order is preserved during time interval $\tau_{SL}$ by introducing spin-locking by a strong RF-field (its amplitude does not match $J$ in order to prevent from spin order conversion). Finally, with the aim to observe the singlet state, the singlet spin order is converted back into transverse magnetization by an RF-field applied during a time period $\tau_{SLIC}$ along the $y$-axis of the rotating frame (again, the RF-amplitude matches the $J$ value).

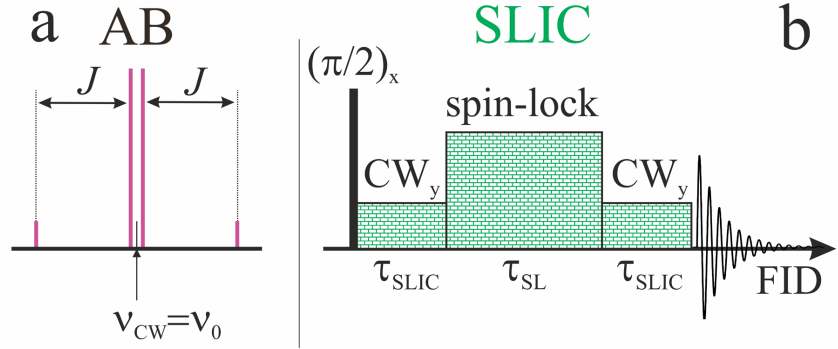

**Figure 1C.** (a) Setting of the RF-frequency for SLIC and (b) experimental protocol used in the SLIC method. In (a) the off-set, Δ, of the RF-frequency from $\nu_0$ is shown, which provide M2S/S2M conversion (for transverse magnetization). The protocol shown in (b) comprises 5 stages. In stage 1 transverse magnetization is formed by a $\left(\frac{\pi}{2}\right)_x$-pulse; then, in stage 2, the magnetization is converted into an LLS the RF-field, which has the $y$-phase and duration of $\tau_{SLIC}$. In stage 3 the singlet states is sustained during a variable time interval $\tau_{SL}$ (by using spin locking or in the absence of spin-locking when $J \gg \delta\nu$). In stage 4 the singlet state is converted back into $y$-magnetization by the RF$_2$-field, which has the $y$-phase and duration of $\tau_{SLIC}$. Finally, in stage 5 the NMR spectrum is taken.

First, we evaluated the spin order conversion efficiency in APSOC and in SLIC by numerical calculations, see **Figure 2C**. For a strongly-coupled spin system both techniques work well; as far as the time required for spin order conversion is concerned, SLIC works better because it provides faster conversion. At the same time, SLIC requires precise setting of the mixing time $\tau_{SLIC}$, whereas in APSOC it is sufficient that $\tau_{on}$ exceeds a certain threshold value. When the spin system is coupled weakly, the SLIC method does not work well anymore. This situation is, in principle, beyond the range of applicability of the method, since there is no crossing at $\nu_1 = |J|$ in a weakly-coupled spin system. However, SLIC can still perform M2S transfer but (i) its efficiency becomes low and (ii) the $\tau_{SLIC}$ dependence contains fast oscillations, so that the transfer process can be difficult to control. At the same time, APSOC works perfectly in both cases.

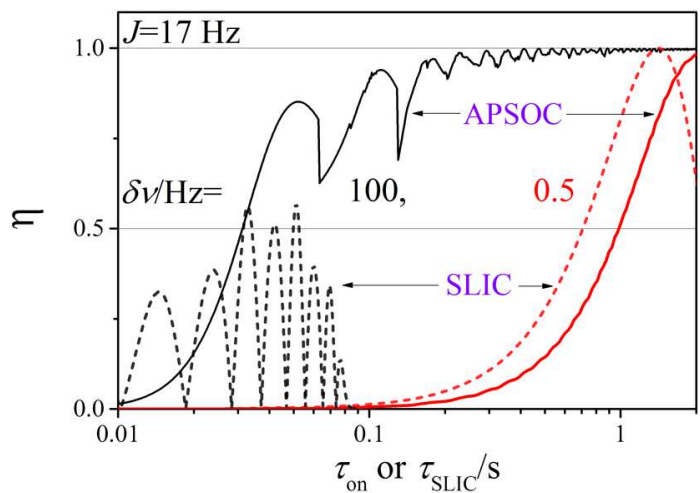

**Figure 2C**. Calculation of the relative performance for the M2S conversion efficiency: dependence of singlet-state population of $\tau_{on}$ for APSOC (solid lines) and dependence of the singlet state population on $\tau_{SLIC}$ for SLIC (dashed lines). The APSOC efficiency is measured as described by Eq. (A6).The calculation is done without taking into account relaxation processes. Here $J = 17$ Hz, $\Delta = 10$ Hz, $\delta\nu$ is 100 Hz (black lines) or 0.5 Hz (red lines). In APSOC the RF-field switching on profiles are linear. The parameters of the field profile (the $\nu_1^{max}$ value) are optimized to obtain the maximal APSOC efficiency for the given spin-system and $\tau_{on}$ and $\Delta$ parameters.

We also performed additional experiments to compare the performance of both techniques. In **Figure 3C** such a comparison is presented for the strongly coupled spin pair of the Gly-residue in the dipeptide. It is clearly seen that both methods perform the conversion and sustain the LLS (NMR signals are visible after 60 s of spin-locking), but APSOC always provides higher NMR signal intensities. This is demonstrated by the $\tau_{SL}$-dependence, see **Figure 3Cc**: in SLIC and APSOC the signals decay at the same rate but in APSOC the signals are always stronger. When $\tau_{SL} \to 0$ the M2S-S2M conversion efficiency for APSOC is $\varepsilon = 0.72$, whereas for SLIC it is below 0.5. By fitting the $\tau_{SL}$ dependence as a sum of two exponents (the slow component stands for the LLS) we have found $A_S = 0.52$, i.e., a strong contribution from the singlet spin order. In a weakly-coupled spin pair (Cys-residue) the performance of APSOC is the same as in the previous case whereas the performance of SLIC is strongly reduced (as expected), see **Figure 4C**.

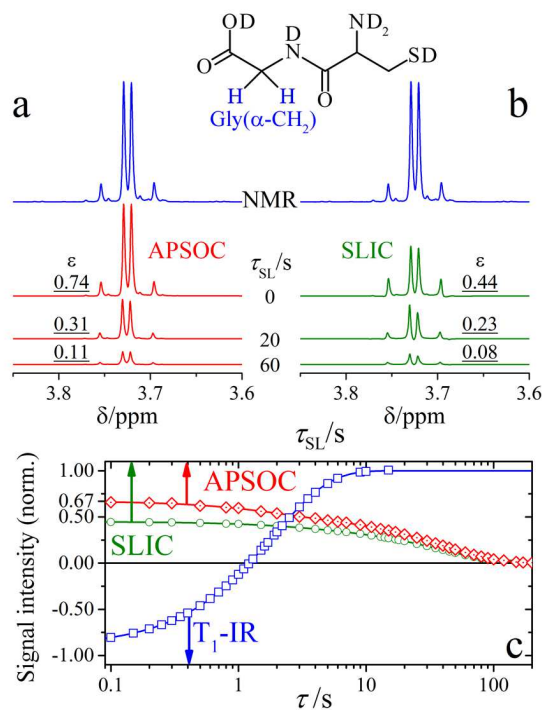

**Figure 3C**. (a) and (b) 700 MHz $^1$H-NMR spectra of Cys-Gly(α-CH$_2$) with APSOC and SLIC spectra for different duration, $\tau_{SL}$, of the spin-locking stage . Signal intensities, $\varepsilon$, are shown at the spectra (underlined numbers). (c) $\tau_{SL}$ time dependence of the NMR intensity of the α-CH$_2$ protons. The signal intensity is given in units of the thermal polarization, indicating that after APSOC we obtain about 3/4 of the starting spin magnetization and in SLIC less than 0.5. The $\tau_{SL}$ dependences are fitted by the function $\varepsilon(\tau_{SL}) = A_T \times \exp(-\tau_{SL}/T_T) + A_S \times \exp(-\tau_{SL}/T_S)$; the fitting parameters for APSOC are: $T_T = 1.73$ s, $T_S = 39.8$ s, $A_T = 0.17$, $A_S = 0.57$ and for SLIC: $T_T = 2.9$ s, $T_S = 39.6$ s, $A_T = 0.055$, $A_S = 0.385$. Experimental APSOC parameters are: $\tau_{on} = \tau_{off} = 0.4$ s, $v_1^{max} = 70$ Hz, $v_{SL} = 1$ kHz, $\tau_{SL}$ is varied, $v_0 = 3.726$ ppm, $\Delta = 12$ Hz. Experimental SLIC parameters are: $\tau_{SLIC} = 45.46$ ms, $v_{SLIC} = 17.5$ Hz, $v_{SL} = 1$ kHz, $\tau_{SL}$ is varied, $v_0 = 3.726$ ppm. In (c) we also show the result of the conventional inversion-recovery experiment (dependence of the NMR signal on the recovery time) used for T$_1$-measurements providing $T_1 = 1.9$ s.

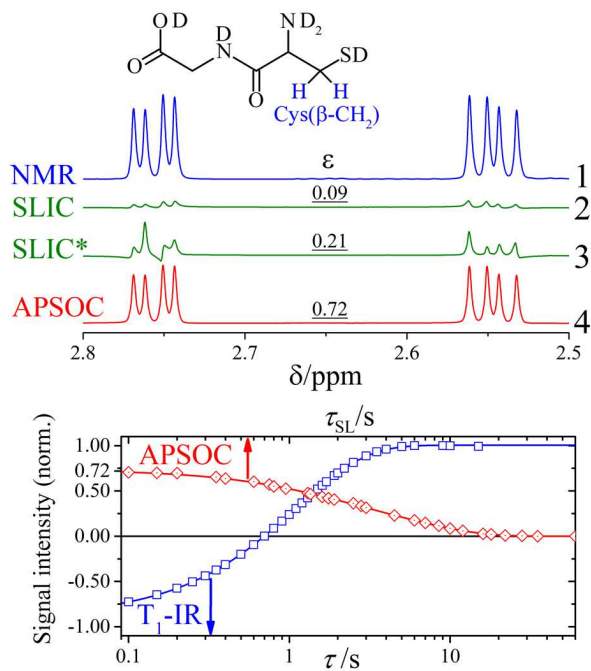

**Figure 4C**. (top) 700 MHz $^1$H-NMR spectrum of the Cys(β-CH$_2$) protons of Cys-Gly (trace 1); SLIC spectrum obtained with parameters $\tau_{SLIC} = 4.8$ ms, $\nu_{SLIC} = 12.7$ Hz, $\tau_{SL} = 0$, $\nu_0 = 2.651$ ppm (trace 2); optimized SLIC* spectrum obtained with parameters $\nu_{SLIC} = 37$ Hz, $\tau_{SL} = 100$ ms and $\nu_0 = 2.651$ ppm (trace 3); APSOC spectrum obtained with parameters: $\tau_{on} = \tau_{off} = 0.2$ s, $\nu_1^{max} = 550$ Hz, $\nu_{SL} = 1$ kHz, $\tau_{SL} = 0$, $\nu_0 = 2.651$ ppm, $\Delta = 10$ Hz (trace 5). Signal intensities, $\varepsilon$, given in units of $M_0$, are shown in the spectra by underlined numbers. (bottom) $\tau_{SL}$ dependence of the NMR intensity of Cys(β-CH$_2$) protons in the APSOC experiment. After APSOC $\varepsilon \approx 0.72$; the $\tau_{SL}$ dependences are fitted by function $\varepsilon(\tau_{SL}) = A_T \times \exp(-\tau_{SL}/T_T) + A_S \times \exp(-\tau_{SL}/T_S)$; the fitting parameters are: $T_T = 1.2$ s, $T_S = 5.8$ s, $A_T = 0.2$, $A_S = 0.52$.

## APPENDIX D. Singlet order maintenance

In our experiments, in order to sustain the long-lived singlet order we use spin locking. The idea behind it is that in the presence of a strong RF-field the singlet state is a true eigen-state of a spin pair. This is a valid assumption once the difference, $\delta\nu_{eff}$, of the precession frequencies of the spins (in the rotating frame) becomes smaller than $J$:

$$|\delta\nu_{eff}| \ll |J| \tag{1D}$$

Let us consider this condition in further detail. In the rotating frame the spins precess at the following frequencies:

$$\nu_a^{eff} = \sqrt{\Delta_a^2 + \nu_1^2}; \quad \nu_b^{eff} = \sqrt{\Delta_b^2 + \nu_1^2} \tag{2D}$$

When $\nu_1 \gg |\Delta_a|, |\Delta_b|$ we obtain that $\delta\nu_{eff} \to 0$, i.e., the singlet state is indeed maintained by the RF-field.

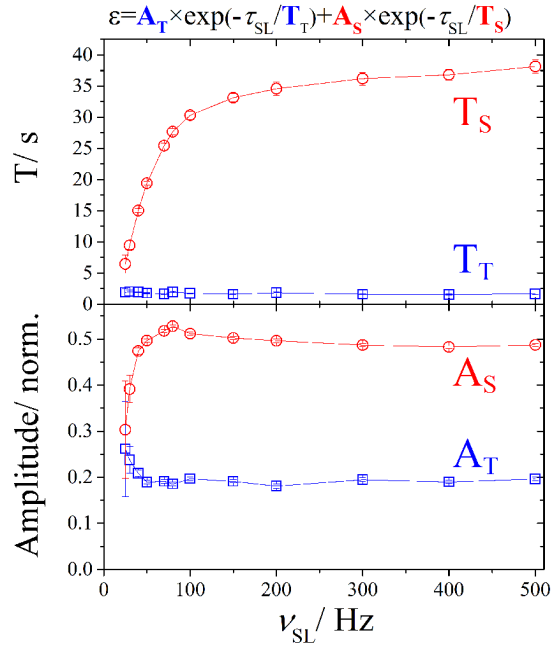

**Figure 1D**. Dependence of the singlet-maintenance efficiency on the amplitude of the spin-locking RF-filed, $\nu_{SL}$, for the lifetimes of the fast, $T_T$, and slow, $T_S$, components (top) and weights of these components (bottom) obtained for the α-CH$_2$ protons of the Gly-residue of Cys-Gly. Experimental parameters are: $\tau_{on} = \tau_{off} = 0.4$ s, $\nu_1^{max} = 70$ Hz, $\nu_0 = 3.726$ ppm, $\Delta = 12$ Hz, $B_0 = 16.4$ T, linear field switching profile is used. To obtain the singlet maintenance efficiency we fitted the $\varepsilon(\tau_{SL})$ dependence by a bi-exponential function. As above, $\varepsilon$ is the spin magnetization obtained in the APSOC experiment measured in units of the thermal magnetization.

A specific case is given by a strongly coupled pair: in such a pair the singlet state is in good approximation an eigen-state with the consequence that it is sustained even without spin-locking, for example, see Ref. [6]. In the example presented here, the Cys-Gly peptide, the LLS maintenance in the absence of spin-locking is not so efficient even for the CH$_2$-protons of Gly ("strongly-coupled" spin pair). To analyze the effect of the spin-locking RF-field we analyzed the $\tau_{SL}$-dependence of NMR signals in APSOC at variable $\nu_{SL}$. The time dependence was fitted by a biexponential function: the fast component was attributed to triplet relaxation with a characteristic time $T_T$ while the slow component was attributed to the LLS. The results are given in

**Figure 1D**: the $T_S$ time is relatively short in the absence of spin-locking and strongly increases upon increase of $\nu_{SL}$. At the same time, the weight of the slow component increases, indicating that spin-locking improves the LLS maintenance.

**References**

[1] M.H. Levitt, Singlet Nuclear Magnetic Resonance, in: M.A. Johnson, T.J. Martinez (Eds.) Ann. Rev. Phys. Chem., Annual Reviews, Palo Alto, 2012, pp. 89-105.
[2] G. Pileio, Relaxation theory of nuclear singlet states in two spin-1/2 systems, Prog. NMR Spectrosc., 56 (2010) 217-231.
[3] A.S. Kiryutin, S.E. Korchak, K.L. Ivanov, A.V. Yurkovskaya, H.-M. Vieth, Creating long-lived spin states at variable magnetic field by means of photo-Chemically Induced Dynamic Nuclear Polarization, J. Phys. Chem. Lett., 3 (2012) 1814-1819.
[4] A.N. Pravdivtsev, A.V. Yurkovskaya, H. Zimmermann, H.-M. Vieth, K.L. Ivanov, Magnetic field dependent long-lived spin states in amino acids and dipeptides, Phys. Chem. Chem. Phys., 16 (2014) 7584-7594.
[5] G. Pileio, M. Carravetta, E. Hughes, M.H. Levitt, The Long-Lived Nuclear Singlet State of 15N-Nitrous Oxide in Solution, J. Am. Chem. Soc., 129 (2008) 12582–12583.
[6] G. Stevanato, J.T. Hill-Cousins, P. Håkansson, S.S. Roy, L.J. Brown, R.C.D. Brown, G. Pileio, M.H. Levitt, A Nuclear Singlet Lifetime of More than One Hour in Room-Temperature Solution, Angew. Chem. Int. Ed., 54 (2015) 3740–3743.
[7] M. Carravetta, O.G. Johannessen, M.H. Levitt, Beyond the $T_1$ Limit: Singlet Nuclear Spin States in Low Magnetic Fields, Phys. Rev. Lett., 92 (2004) 153003.
[8] A. Bornet, P. Ahuja, R. Sarkar, L. Fernandes, S. Hadji, S.Y. Lee, A. Haririnia, D. Fushman, G. Bodenhausen, P.R. Vasos, Long-Lived States to Monitor Protein Unfolding by Proton NMR, ChemPhysChem, 12 (2011) 2729-2734.
[9] R. Sarkar, P.R. Vasos, G. Bodenhausen, Singlet-State Exchange NMR Spectroscopy for the Study of Very Slow Dynamic Processes, J. Am. Chem. Soc., 129 (2007) 328-334.
[10] P. Ahuja, R. Sarkar, P.R. Vasos, G. Bodenhausen, Diffusion Coefficients of Biomolecules Using Long-Lived Spin States, J. Am. Chem. Soc., 131 (2009) 7498-7499.
[11] R. Sarkar, P. Ahuja, P.R. Vasos, G. Bodenhausen, Measurement of Slow Diffusion Coefficients of Molecules with Arbitrary Scalar Couplings via Long-Lived Spin States, ChemPhysChem, 9 (2008) 2414-2419.
[12] J.-N. Dumez, J.T. Hill-Cousins, R.C.D. Brown, G. Pileio, Long-lived localization in magnetic resonance imaging, J. Magn. Reson., 246 (2014) 27-30.
[13] G. Pileio, J.-N. Dumez, I.-A. Pop, J.T. Hill-Cousins, R.C.D. Brown, Real-space imaging of macroscopic diffusion and slow flow by singlet tagging MRI, J. Magn. Reson., 252 (2015) 130-134.
[14] S. Cavaldini, P.R. Vasos, Singlet States Open the Way to Longer Time-Scales in the Measurement of Diffusion by NMR Spectroscopy, Conc. Magn. Reson., 32A (2008) 68-78.
[15] R. Buratto, D. Mammoli, E. Chiarparin, G. Williams, G. Bodenhausen, Exploring Weak Ligand–Protein Interactions by Long-Lived NMR States: Improved Contrast in Fragment-Based Drug Screening, Angew. Chem. Int. Ed., 53 (2014) 11376-11380.
[16] A. Bornet, S. Jannin, G. Bodenhausen, Three-field NMR to preserve hyperpolarized proton magnetization as long-lived states in moderate magnetic fields, Chem. Phys. Lett., 512 (2011) 151-154.
[17] P.R. Vasos, A. Comment, R. Sarkar, P. Ahuja, S. Jannin, J.-P. Ansermet, J.A. Konter, P. Hautle, B. van den Brandt, G. Bodenhausen, Long-lived states to sustain hyperpolarized magnetization, Proc. Natl. Acad. Sci. USA, 106 (2009) 18469-18473.
[18] G. Pileio, S. Bowen, C. Laustsen, M.C.D. Tayler, J.T. Hill-Cousins, L.J. Brown, R.C.D. Brown, J.H. Ardenkjaer-Larsen, M.H. Levitt, Recycling and Imaging of Nuclear Singlet Hyperpolarization, J. Am. Chem. Soc., 135 (2013) 5084-5088.
[19] M.C.D. Tayler, I. Marco-Rius, M.I. Kettunen, K.M. Brindle, M.H. Levitt, G. Pileio, Direct Enhancement of Nuclear Singlet Order by Dynamic Nuclear Polarization, J. Am. Chem. Soc., 134 (2012) 7668-7671.
[20] A.S. Kiryutin, H. Zimmermann, A.V. Yurkovskaya, H.-M. Vieth, K.L. Ivanov, Long-lived spin states as a source of contrast in magnetic resonance spectroscopy and imaging, J. Magn. Reson., 261 (2015) 64-72.
[21] A.S. Kiryutin, K.L. Ivanov, A.V. Yurkovskaya, H.-M. Vieth, N.N. Lukzen, Manipulating spin hyper-polarization by means of adiabatic switching of RF-field, Phys. Chem. Chem. Phys., 15 (2013) 14248-14255.

[22] M.H. Levitt, Symmetry Constraints on Spin Dynamics: Application to Hyperpolarized NMR, J. Magn. Reson., 262 (2015) 91-99.

[23] Software for optimizing parameters for APSOC, http://www.tomo.nsc.ru/en/nmr/sos-filter/.

[24] S.J. DeVience, R.L. Walsworth, M.S. Rosen, Preparation of Nuclear Spin Singlet States Using Spin-Lock Induced Crossing, Phys. Rev. Lett., 111 (2013) 173002.

[25] G. Pileio, M. Carravetta, M.H. Levitt, Storage of nuclear magnetization as long-lived singlet order in low magnetic field, Proc. Natl. Acad. Sci. USA, 107 (2010) 17135-17139.

[26] J. Natterer, J. Bargon, Parahydrogen induced polarization, Prog. Nucl. Magn. Reson. Spectrosc., 31 (1997) 293-315.

[27] R.A. Green, R.W. Adams, S.B. Duckett, R.E. Mewis, D.C. Williamson, G.G.R. Green, The theory and practice of hyperpolarization in magnetic resonance using parahydrogen, Prog. Nucl. Magn. Reson. Spectrosc., 67 (2012) 1-48.

[28] N.A. Gershenfeld, I.L. Chuang, Bulk spin-resonance quantum computation, Science, 275 (1997) 350-356.

[29] M.A. Nielsen, I.L. Chuang, Quantum Computation and Quantum Information, Cambridge University Press, Cambridge, UK, 2002.